\documentclass[preprint,aps,pra,showkeys,showpacs]{revtex4-1}
\usepackage{graphicx}
\begin{document}

\title{Thickness and strain effects on the thermoelectric transport in nanostructured Bi$_2$Se$_3$}

\author{Y. Saeed, N. Singh, U. Schwingenschl\"ogl}

\email{udo.schwingenschlogl@kaust.edu.sa,+966(0)544700080}

\affiliation{KAUST, PSE Division, Thuwal 23955-6900, Kingdom of Saudi Arabia}

\begin{abstract}
The structural stability, electronic structure, and thermal transport properties
of one to six quintuple layers (QLs) of Bi$_2$Se$_3$ are investigated by van 
der Waals density functional theory and semi-classical Boltzmann theory. 
The bandgap amounts to 0.41 eV for a single QL and reduces to 0.23 eV when
the number of QLs increases to six. A single QL has a significantly higher thermoelectric
figure of merit (0.27) than the bulk material (0.10), which can be further enhanced to 0.30
by introducing 2.5\% compressive strain. Positive phonon frequencies under strain indicate
that the structural stability is maintained.
\end{abstract}

\maketitle

Thermoelectric (TE) materials have been extensively studied for key applications such
as the conversion of waste heat into electricity. They have huge potential to contribute
to a solution of the energy and environment crisis. The efficiency of TE devices is
determined by the dimensionless figure of merit $ZT={\sigma}S^{2}T/\kappa$, where $\sigma$ 
is the electrical conductivity, $S$ is the Seebeck coefficient, $T$ is the temperature, 
and $\kappa$ is the thermal conductivity. The latter comprises lattice
($\kappa_l$) and electronic ($\kappa_e$) contributions: $\kappa=\kappa_l+\kappa_e$. 
In order to achieve a high efficiency, the TE material has to be a good electrical and
poor thermal conductor and, at the same time, has to possess a high Seebeck coefficient,
see, for example, Ref.\ \cite{Bell} and the references therein. Nanostructuring, alloying,
and the application of strain are effective and frequently used approaches
\cite{Venkat,Harman,Soni} to improve TE properties, because the phonon scattering is
enhanced and the thermal conductivity consequently is reduced.

The class of topological insulators has been proposed to be of particular potential for
TE materials \cite{Snyder}. A well known example is bismuth telluride (Bi$_2$Te$_3$),
which shows a high figure of merit of about 1 \cite{tritt}, but also many other members
of this family of materials, such as Bi$_2$Se$_3$ and Sb$_2$Te$_3$, are topological
insulators with an energy gap in the bulk and gapless surface states. The high mobility of
the surface electrons combined with the bulk energy gap give rise to the superior
thermoelectric properties. In particular, superlattices of Te-based materials
(Bi$_2$Te$_3$/Sb$_2$Te$_3$ \cite{Venkat}, PbSe$_{0.98}$Te$_{0.02}$/PbTe \cite{Harman})
and nanocrystalline Bi$_x$Sb$_{2-x}$Te$_3$ \cite{Poudel} have been reported to exhibit 
a high $ZT$. However, the best performing materials contain Te, which is an expensive
element and of limited availability. Therefore, alternative materials (for example based
on Se) are required. Bi$_2$Se$_3$ is a topological insulator with an energy gap of 0.3 eV
in the bulk. It has a rhombohedral crystal structure with lattice parameters of $a=4.143$
\AA\ and $c=28.636$ \AA\ (space group $R\bar{3}m$) \cite{Nakajima}. The layered structure
is based on slabs of five shifted Se and Bi atomic layers stacked along the $c$-axis with
the sequence Se(1)-Bi-Se(2)-Bi-Se(1). Such a slab is called quintuple layer (QL) and
has a thickness of about 1 nm. Along the $c$-axis the QLs are connected by weak van der
Waals (vdW) forces. Although the vdW interaction is relatively weak, it still plays a
crucial role for the electronic and TE properties.  

Various approaches have been proposed to improve the TE properties of bulk Bi$_2$Se$_3$,
such as the introduction of point defects and doping of Sb, Cu, or Ca atoms to tune the
carrier concentration \cite{Xue-pointdef,Kong-sb,Martin-Cu,Hor-Ca}. Recent reports
show that mechanical exfoliation of a single QL and of slabs of a few QLs of Bi$_2$Se$_3$
is possible and that excellent TE properties can be expected in such thin films
\cite{Rao,Sun,Sakamoto1238ql,zhang-bise-ql,Zhang-QL-apl}. Sun and coworkers \cite{Sun} have
shown that the TE behavior of a single QL of Bi$_2$Se$_3$ is enhanced because of a reduced
$\kappa$, as compared to the bulk value. When the number of QLs increases the bandgap of
Bi$_2$Se$_3$ decreases and is finally closed at a thickness of eight QLs \cite{Sakamoto1238ql}.
This strong variation of the bandgap with the thickness will play a vital role for the
tunability of the TE properties. In this context, we investigate the effect of the thickness
in terms of the number of QLs on the electronic structure and transport behavior.
We will demonstrate a high figure of merit for a single QL of Bi$_2$Se$_3$ and will argue
that the value can be further enhanced by tensile strain (for which we show that the
material remains stable). 

We have carried out first principles calculations based on density functional theory,
employing the generalized gradient approximation of the exchange correlation potential
in the Perdew-Burke-Ernzerhof flavor. The vdW interaction is taken into account by
means of the B97-D functional \cite{Grimme} as implemented in the QuantumESPRESSO code
\cite{QE}. All calculations are performed with a
plane wave energy cutoff of 544 eV and a Monkhorst-Pack $8\times8\times1$ k-mesh for
the Brillouin zone integrations. Furthermore, we employ finer $16\times16\times1$ and
$40\times40\times1$ k-meshes for calculating the density of states (DOS) and transport
properties, respectively. All structures are optimized until an energy convergence of
10$^{-5}$ eV and a force convergence of 0.025 eV/\AA\
are reached. A 15 \AA\ vacuum slab is used to prevent artificial interaction with
periodic images due to the employed periodic boundary conditions. The phonon frequencies
are determined by density functional perturbation theory for evaluating the structural
stability \cite{Baroni}.

Once self-consistency is achieved in the first principles calculations, the transport
properties are calculated using the semi-classical Boltzmann theory within the constant
relaxation time approximation, as implemented in the BoltzTraP code \cite{boltz}. 
This approach yields accurate results for various types of TE materials \cite{saeed2,2,3,4,amin}.
The transport function is given by
\begin{equation}
\sigma(E) = DOS(E)v{^2}(E)\tau(E),
\end{equation}
the electrical conductivity by
\begin{equation}
\sigma(T) = -\int_{-\infty}^{\infty}dE\sigma(E)\frac{df(E-\mu)}{dE},
\end{equation}
and the Seebeck coefficient by
\begin{equation}
S(T) = -\frac{k_B}{e\sigma(T)}\int_{-\infty}^{\infty}dE\sigma(E)\frac{E-\mu}{T}\frac{df(E-\mu)}{dE},
\end{equation}
where $f$ is the Fermi function, $\tau(E)$ the scattering time, $v(E)$ the Fermi velocity, and
$\mu$ the chemical potential.
 
In order to built supercells for one to six QLs of Bi$_2$Se$_3$, we start from a single 
QL using the bulk structural parameters and optimize the in-plane lattice constant,
obtaining a slight elongation to $a=4.191$ \AA. The thickness of the QL increases from the
experimental bulk value of 6.966 \AA\ \cite{Nakajima} to 7.096 \AA, similar to the results
in Ref.\ \cite{Liu-prb}. We then keep the $a$ lattice constant for all systems
fixed at the value obtained for the single QL and relax in each case the atomic positions.
The resulting values for the layer thickness ($h$) and separation between adjacent QLs ($d$)
are summarized in Fig.\ 1. While effects of the vdW correction expectedly are negligible 
for a single QL, our values for $h$ and $d$ for two and more QLs are larger than previously
reported \cite{Liu-prb}. We note that we use an optimized lattice constant, while
these authors have employed the bulk value. The effect of the vdW correction is significant
and cannot be ignored in the case of two and more QLs in order to obtain correct results
for $h$ and $d$. As a consequence, our calculated energy gaps better reproduce the experiment
\cite{Sakamoto1238ql} than Ref.\ \cite{Liu-prb}, see the band structures of the relaxed
systems in Fig.\ 2. This improved agreement is due to the larger $d\sim3.3$ \AA\ (as compared
to $\sim3.0$ \AA). Our results show that the single QL has a bandgap of 0.41 eV, which
decreases step by step to 0.23 eV for six QLs, see Fig.\ 3. The largest Seebeck coefficient
and figure of merit thus are expected for a single QL of Bi$_2$Se$_3$.

We next apply both compressive and tensile strains of 2.5\% and 5\% to explore the effects
on the TE response. In the single QL case the layer thickness is reduced to 6.725 \AA\ and
6.920 \AA\ for 5\% and 2.5\% tensile strain, respectively, and grows to 7.470 \AA\ and
7.280 \AA\ for 5\% and 2.5\% compressive strain. Particularly, at 2.5\% tensile strain the
value is almost equal to the bulk lattice constant. The band structures of a single QL of
Bi$_2$Se$_3$ under strain are depicted in Fig.\ 4 together with the corresponding phonon
dispersions: The bandgap reaches 0.60 eV for 5\% compressive strain and decreases to
0.16 eV for 5\% tensile strain. 2.5\% tensile strain applied to a single QL of Bi$_2$Se$_3$
has about the same effect as an increase of the thickness to 4 QLs. This is just an example
to demosntrate that the bandgap can be tune almost idependently by the strain and the
number of QLs, which is a great advantage from the technology point of view. 

We obtain positive phonon frequencies in the whole range of strain under study for the
single QL, which demonstrates a wide stability window. There are 15 phonon modes at 
the center of the Brillouin zone, 12 optical and 3 acoustical. The optical modes consist
of A$^1_{1g}$ and A$^2_{1g}$ out-of-plane modes and E$^1_g$ and E$^2_g$ in-plane modes. 
Nanostructured Bi$_2$Se$_3$ is subject to a red shift of the frequencies of the above
modes due to phonon softening. The out-of-plane phonon softening is explained by the
finite size along the $c$-axis, resulting in a reduction of the restoring force. 
In an unstrained single QL of Bi$_2$Se$_3$, the E$^2_g$ and A$^2_{1g}$ frequencies are
129.3 cm$^{-1}$ and 174.8 cm$^{-1}$, respectively, and thus close to the experimental
bulk values of 131 cm$^{-1}$ and 174 cm$^{-1}$ \cite{Zhang-phonon}. Strain effects on
the A$^2_{1g}$ frequency are addressed in Fig.\ 5 in comparison to the variation of the
bandgap, in order to demonstrate the analogous behavior of these two quantities.
Under tensile strain the frequency decreases, i.e., the out-of-plane phonons soften,
while in the case of compressive strain it increases. 

The Gr\"uneisen parameter for the A$^2_{1g}$ mode (i.e., the G mode at the $\Gamma$ point)
is given by $\gamma_G=-\Delta\omega_G/2\omega_G^0\varepsilon$, where $\Delta\omega_G$ is
the difference of the frequencies with and without strain and $\omega_G^0$ is the
frequency of the unstrained single QL. We obtain positive values of 1.2 and 2.6 for
2.5\% and 5\% tensile strain, respectively, because the distance between the Bi and Se
sublattices is reduced. Accordingly, negative values appear for compressive strain.
So far no experimental data are available for comparison, though an experimental confirmation
of our predictions by Raman spectroscopy would be desirable.

We now turn to the TE properties of nanostructured Bi$_2$Se$_3$ with varying thickness.
$S$ and $S^2\sigma/\tau$ are shown in Fig.\ 6 as functions of the hole doping $\rho$
(in cm$^{-3}$). The highest Seebeck coefficient is found for the single QL around
$\rho=2.2\times10^{17}$ cm$^{-3}$. At low doping the value progressively reduces when
the number of QLs grows, because the bandgap decreases. Above $\rho=1.2\times10^{18}$
cm$^{-3}$ the Seebeck coefficient decreases in each case for increasing hole doping.
The reasons for the particularly high values found for the single QL are (1) the maximal
bandgap and (2) the qualitatively different shape of the valence bands.

The BoltzTraP code obtains the electrical conductivity as $\sigma/\tau$. We estimate
$\tau$ for Bi$_2$Se$_3$ by comparing the calculated values of $\sigma/\tau$ and $S$ 
with the experimental values of $S$ and $\sigma$ at 300 K, which results in 
$\tau=2.7\cdot10^{-15}$ s. This value is used for all subsequent calculations of the TE
properties. A similar methodology has been used to establish the relaxation time
in Bi$_2$Te$_3$ \cite{tau} and has led to reasonable agreement with the experiment.
The assumption that $\tau$ is independent of both the thickness and the strain is supported
by previous theoretical work on Sb$_2$Te$_3$ \cite{Thonhauser}. Experimentally, a
thermal conductivity of $\kappa=0.49$ W/Km at a temperature of 300 K has been reported
for nanostructured Bi$_2$Se$_3$ in Ref.\ \cite{Sun} and is used in the following.

According to Fig.\ 6, the maximum of $S^2\sigma/\tau$ for the single QL is found at
a hole doping of $\rho=2.4\times10^{20}$ cm$^{-3}$, which is below the degenerate limit.
It shifts to higher $\rho$ when the number of QLs grows, except for the case of 2 QLs.
Can the value be further improved by strain? We find that the maximal Seebeck coefficient
shifts to higher and lower hole doping, respectively, for tensile and compressive strain.
While $S^2\sigma/\tau$ is reduced by tensile strain as well as by strong compressive
strain, a moderate compressive strain results in a significant enhancement around
$\rho=10^{20}$ cm$^{-3}$. Using the experimental $\kappa=0.49$ W/Km and calculated
$\tau=2.7\cdot10^{-15}$ s we estimate a figure of merit of 0.27 for the single QL, which
is remarkably more than the bulk value (0.10). A moderate compressive strain (2.5\%)
further enhances the value to 0.30. In addition, it can be expected that $S$ and thus
the figure of merit grow with the temperature and that a decreasing $\tau$
for higher doping also has positive effects.

In conclusion, we have established the dependence of the TE properties of nanostructured
Bi$_2$Se$_3$ on the thickness of the material and on applied compressive and tensile
strain by density functional theory and semi-classical Boltzmann theory. We have
demonstrated that a careful description of structural properties, mainly the 
thickness of the QLs and their separation, as determined by the van der Waals interaction,
are essential for obtaining quantitatively correct bandgaps. We have shown that the bandgap
decreases when the number of QLs increases, which results in smaller Seebeck coefficients.
The bandgap of a single QL of Bi$_2$Se$_3$ reaches a value of 0.6 eV for 5\% compressive
strain, whereas we obtain 0.16 eV for 5\% tensile strain. We have confirmed the structural
stability of the systems under strain by means of phonon dispersions. The highest
figure of merit is found in the thinnest system (single QL) and our results demonstrate
that it is further enhanced by compressive strain. A value of 0.30 is achieved for a feasible
strain of 2.5\%, which is three times the bulk value. Strain engineering of 
nanostructured Bi$_2$Se$_3$ therefore turns out to be a highly efficient approach for
creating high performance TE materials. 

\section*{Acknowledgement} We thank KAUST research computing for supplying the computational
resources for this study. N.S.\ acknowledges SABIC for financial support.

\clearpage

\begin{figure}[t]
\includegraphics[width=0.6\textwidth,clip]{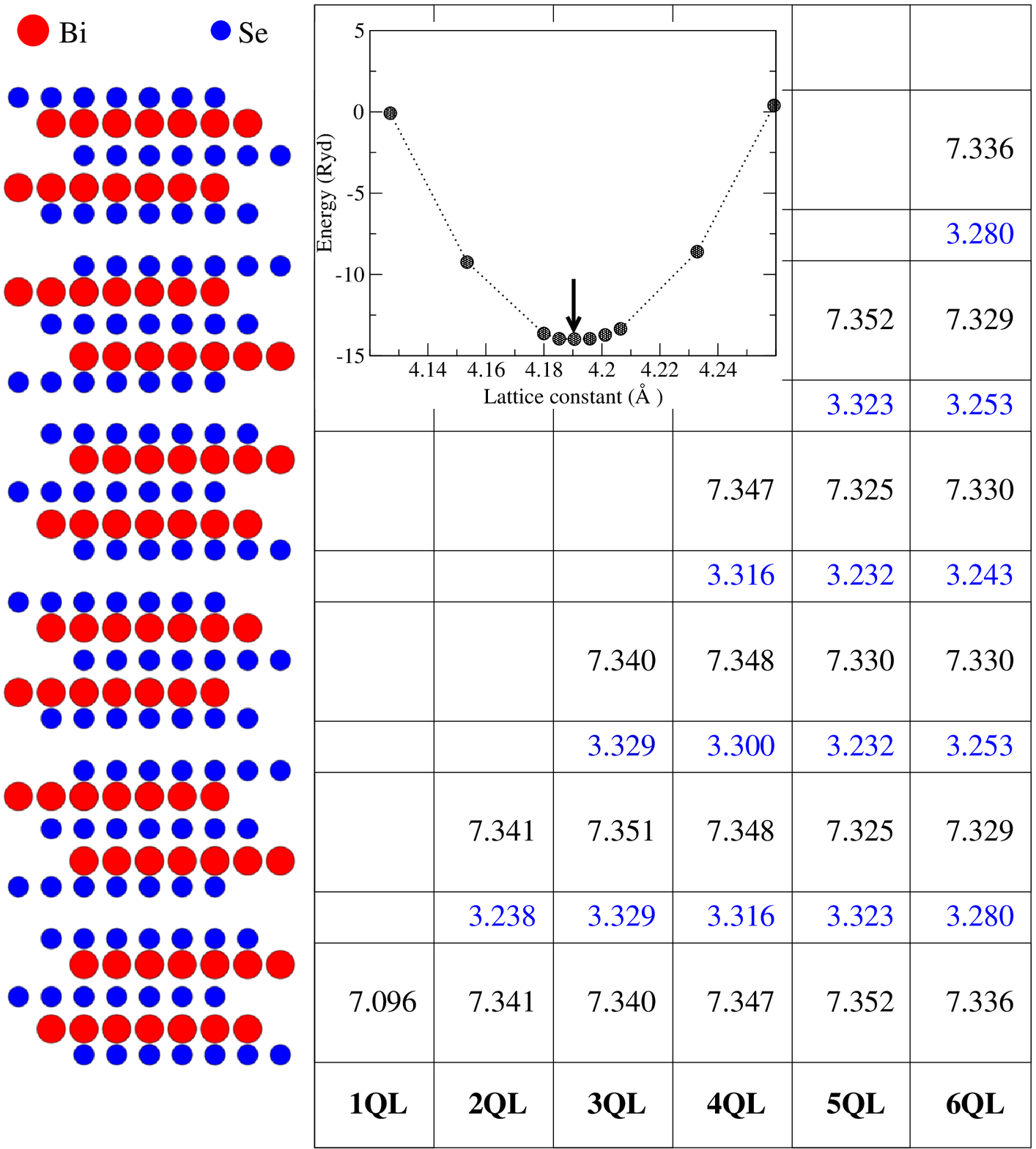}
\caption{Structures of one to six QLs of Bi$_2$Se$_3$ after relaxation: thickness $h$ (black
numbers) and separation $d$ (blue numbers). All values are given in \AA.
The inset addresses the optimization of the lattice parameter for a single QL.}
\end{figure}

\begin{figure}[t]
\includegraphics[width=1.0\textwidth,clip]{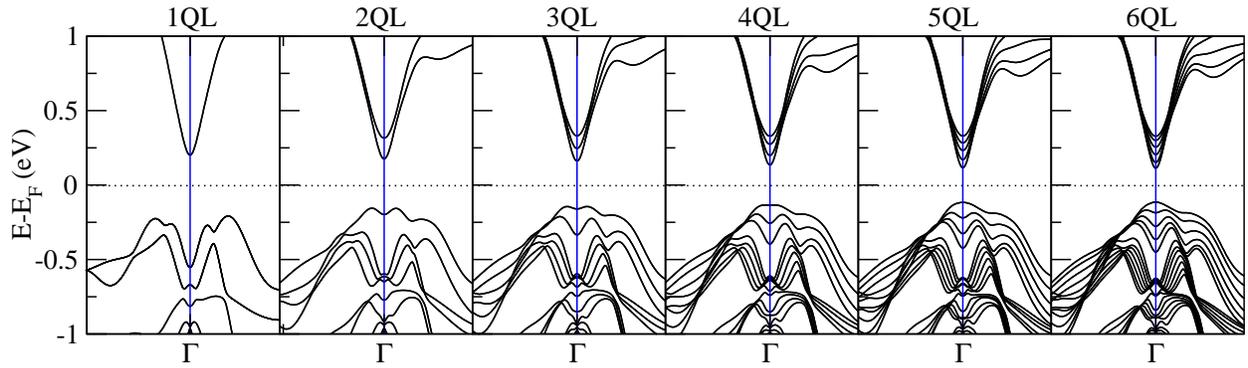}
\caption{Band structures of one to six QLs of Bi$_2$Se$_3$.}
\end{figure}

\begin{figure}[t]
\includegraphics[width=0.5\textwidth,clip]{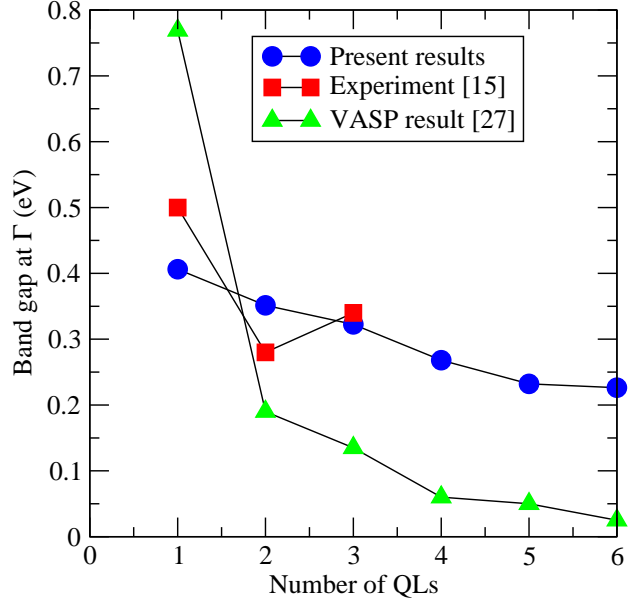}
\caption{Bandgap as a function of the number of QLs of Bi$_2$Se$_3$, along
with previous experimental and theoretical results.}
\end{figure}

\begin{figure}[t]
\includegraphics[width=0.6\textwidth,clip]{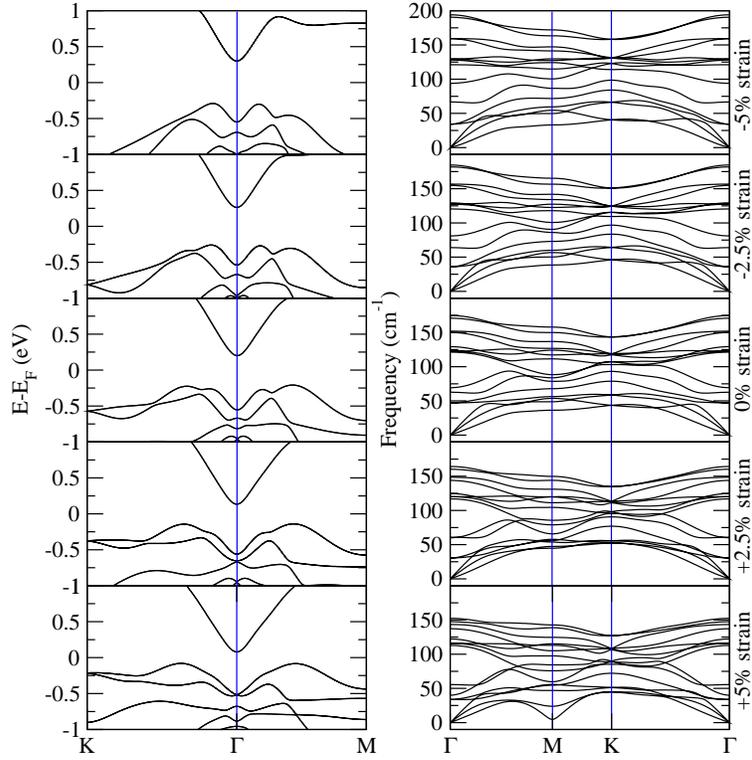}
\caption{Electronic band structures and phonon dispersions of a single QL of Bi$_2$Se$_3$
for different strains.}
\end{figure}

\begin{figure}[t]
\includegraphics[width=0.5\textwidth,clip]{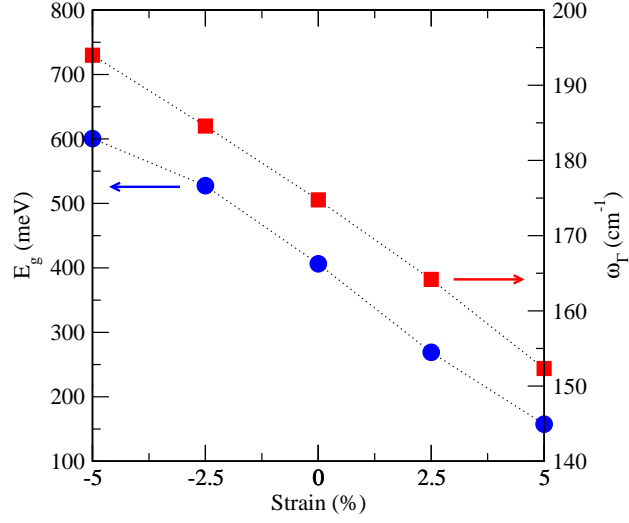}
\caption{Bandgap and $\Gamma$ point phonon frequency of a single QL of Bi$_2$Se$_3$.}
\end{figure}

\begin{figure}[t]
\includegraphics[width=1.0\textwidth,clip]{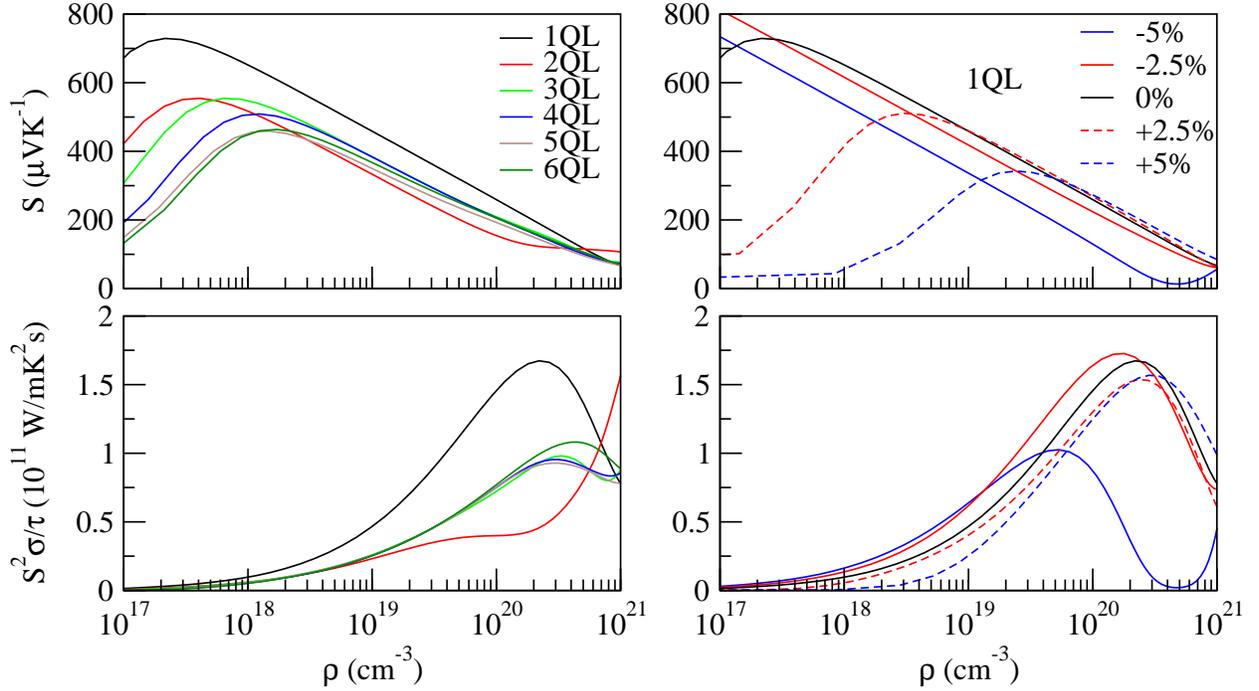}
\caption{Thermoelectric properties for systems of different thickness
and for various strains at 300 K.}
\end{figure}

\end{document}